\documentclass[aps,preprintnumbers,superscriptaddress,runinaddress,amsmath,amssymb,showpacs,floatfix,twocolumn]{revtex4-1}

\usepackage{graphicx}
\usepackage{amsmath}

\DeclareMathOperator{\Tr}{Tr}

\newcommand{\dt}{dt}

\begin{document}

\title{Search for low-lying lattice QCD eigenstates in the Roper regime}

%=========================================

\author{Adrian L. Kiratidis}
\email{adrian.kiratidis@adelaide.edu.au} 
\author{Waseem Kamleh} 
\author{Derek B. Leinweber}
\author{Zhan-Wei Liu} 
\author{Finn M. Stokes}
\author{Anthony W. Thomas}
  \affiliation{Special Research Centre for the Subatomic Structure of
  Matter, Department of Physics, University of Adelaide,
  South Australia 5005, Australia.}

\begin{abstract}
The positive-parity nucleon spectrum is explored in $2 + 1$-flavour lattice QCD in a search for 
new low-lying energy eigenstates near the energy regime of the Roper resonance.
In addition to conventional three-quark operators, we consider novel, local five-quark meson-baryon
type interpolating fields that hold the promise to reveal new eigenstates that may have been missed
in previous analyses.
Drawing on phenomenological insight, five-quark operators based on $\sigma{N}$, $\pi{N}$ and
$a_0{N}$ channels are constructed.
Spectra are produced in a high-statistics analysis on the PACS-CS dynamical gauge-field
configurations with $m_{\pi} = 411\textrm{ MeV}$ via variational analyses of several operator
combinations.
Despite the introduction of qualitatively different interpolating fields, no new states are
observed in the energy regime of the Roper resonance.
This result provides further evidence that the low-lying finite-volume scattering states are not
localised, and strengthens the interpretation of the Roper as a coupled-channel,
dynamically-generated meson-baryon resonance.
\end{abstract}

\pacs{11.15.Ha,12.38.-t,12.38.Gc}

\maketitle

\section{Introduction}
\label{sect:Introduction}
Since the inception of lattice QCD, significant effort has been invested in exploring hadronic spectra, both to shed light upon the nature and properties of
various states, and to test the validity of the methodology itself.  In particular, the community
has shown notable interest in the positive-parity nucleon
channel~\cite{Mahbub:2010rm,Edwards:2011jj,Roberts:2011ym,Bauer:2012at,Roberts:2013ipa,Liu:2014jua},
where the first positive-parity $J^P = {\frac{1}{2}}^{+}$ excitation of the nucleon, known as the
Roper resonance $N^{*}(1440)$, remains a puzzle.  

In Nature, the Roper lies $95\textrm{ MeV}$ below the lowest-lying negative-parity resonant state,
the $N^{*}(1535)$, whereas in constituent quark models, where it is associated with an $N=2$ radial
excitation of the nucleon, the order is reversed~\cite{Isgur:1977ef,Isgur:1978wd, Glozman:1995fu}.  Speculation about the
true nature of this state has been widespread, including the idea that the Roper can be
understood with five-quark meson-baryon dynamics~\cite{Speth:2000zf}.

A critical challenge for lattice spectroscopy in this channel is to judiciously choose an
appropriate operator basis to sufficiently span states of interest in the low-lying spectrum.  This
can be achieved in multiple ways.  Recalling that any radial function can be expanded using a basis
of different width Gaussians, $f(|\vec{r}|) = \sum_i c_i\, e^{-\varepsilon_i r^2}$, suggests the use
of fermion sources with varying Gaussian smeared widths~\cite{Burch:2004he}, which is one method to
obtain a basis of operators possessing enhanced overlap with radial excitations.

The CSSM lattice collaboration was the first to demonstrate that the inclusion of Gaussian-smeared
fermion sources of wide widths are critical in enabling the extraction of the lowest-lying
positive-parity excitation in the nucleon channel~\cite{Mahbub:2009aa,Mahbub:2010rm}.  It was later
shown that the quark probability distribution for this state is consistent with an $N=2$ radial
excitation~\cite{Roberts:2011ym,Roberts:2013oea}.

Another method for selecting an appropriate operator basis is to include qualitatively different
operators, by introducing interpolating fields with the same quantum numbers but different quark
and/or Dirac structure.  Here, it becomes instructive to briefly examine the contemporary work done
in the negative-parity nucleon channel with its two low-lying resonances, the $N^{*}(1535)$ and
$N^{*}(1650)$~\cite{Bruns:2010sv, Edwards:2011jj, Lang:2012db, Mahbub:2012ri, Mahbub:2013bba}.

In recent years the CSSM and Hadron Spectrum lattice collaborations have studied the low-lying
negative-parity spectrum of the nucleon using various local three-quark operators
\cite{Mahbub:2012ri, Mahbub:2013bba, Mahbub:2013ala, Edwards:2012fx, Edwards:2011jj, Bulava:2010yg}
but were unable to extract a state consistent with the low-lying S-wave $\pi{N}$ scattering
threshold.  
Notably, at near physical quark masses this threshold lies below the lowest-lying
negative-parity resonant state, making it an intuitive place to search for the presence of states
consistent with scattering thresholds.

However, for weakly interacting two-particle states, the probability of finding the second particle
at the position of the first is proportional to $1/V$, where $V$ is the spatial volume of the
lattice.  Therefore, the coupling of weakly interacting scattering states to local operators is
volume suppressed.

Naturally, one would expect five-quark operators to possess higher overlap with five-quark states,
and as such the CSSM lattice collaboration introduced local five-quark $\pi{N}$-type operators
\cite{Kiratidis:2015vpa}.  These operators were constructed from a negative-parity pion piece
together with a positive-parity nucleon piece.  Consequently, the operators were expected to possess
higher overlap with the S-wave $\pi{N}$ scattering state and, indeed, a state consistent with this
threshold was observed.  
However, the coupling was relatively weak, and one can conclude that the S-wave $\pi{N}$ scattering
state is poorly localised and better treated with an approach in which the momenta of both the pion
and the nucleon are projected to zero.  These non-local operators are known to have excellent overlap
with the scattering state \cite{Lang:2012db}.

Turning to the positive-parity channel, we are searching for new states that
have poor overlap with conventional three-quark operators and therefore have been missed in
analyses to date.
Meson-baryon states having strong attraction, which can
give rise to localization of the state \cite{Liu:2016uzk}, are expected to have good overlap with our local
five-quark operators.
The existence of such states would suggest an important role for molecular meson-baryon 
configurations~\cite{Hall:2014uca} in the formation of the Roper resonance.

To obtain positive parity in a local meson-baryon interpolating field, the intrinsic parities of the
meson and baryon must match, and there are two approaches one can consider.
Because the lowest-lying five-quark scattering state is a $\pi{N}$ $P$-wave state, previous attempts
have considered the approach of local $\pi{N}$-type interpolators.  As the ability to construct a
relative $P$-wave $\pi{N}$ doesn't exist in a local operator, this approach necessarily draws on an
odd-parity excitation of the nucleon to form the quantum numbers of the Roper.
As one might expect, this operator had negligible overlap with the $P$-wave $\pi{N}$ scattering
threshold which lies between the ground state and the first positive-parity excitation observed in
lattice QCD at light quark masses.
No state consistent with this threshold was observed in the five-quark analysis of
Ref.~\cite{Kiratidis:2015vpa}.

Drawing on the success of the local $\pi{N}$-type operator in the negative parity sector, we consider the
alternative approach of pairing an even-parity meson interpolator with the nucleon interpolator such
that the ground state nucleon can participate in forming the positive-parity quantum numbers of the
Roper resonance.
In this analysis, we construct the local five-quark meson-baryon operators $a_{0}{N}$ and $\sigma{N}$,
and investigate their impact on the positive-parity nucleon spectrum.  We search for both new
low-lying eigenstates in the finite volume of the lattice, and/or an alteration of the spectrum
reported in previous analyses.

Following the outline of variational analysis techniques in
Section~\ref{sect:CorrelationMatrixTechniques}, we construct the new local five-quark operators in
Section~\ref{sect:InterpolatingFields}, and outline the stochastic-noise methods employed to
calculate the corresponding loop containing diagrams.  Simulation details are discussed in
Section~\ref{sect:Simulation Details} and the results of the variational analyses are presented in
Section~\ref{sect:Results}.  A summary of our findings and their impact our our understanding of
the Roper resonance is presented in Sec.~\ref{sect:Conclusions}.
%    
%%%%%%%%%%%%%%%%%%%%%%%%%%%%%%%%%%%%%%%%%%%%%%
%
\section{Correlation Matrix Techniques}
\label{sect:CorrelationMatrixTechniques}

Correlation matrix based variational analyses~\cite{Michael:1985ne, Luscher:1990ck} are now
well-established as a framework within which hadron spectra can be produced
\cite{Leinweber:2015kyz}. The methodology begins via the judicious selection of a suitably large
basis of $N$ operators, such that the states of interest within the spectrum are contained within
the span.  An $N \times N$ matrix of cross correlation functions,
\begin{equation}
\label{defn:CM}
\mathcal{G}_{ij}(\vec{p},t) = \sum_{\vec{x}}\textrm{e}^{-i\vec{p}\cdot\vec{x}}\,\big\langle\,\Omega\, \big| \,\chi_{i}(\vec{x},t)\,\overline{\chi}_{j}(\vec{0},t_{src})\, \big| \,\Omega\, \big\rangle,
\end{equation}
is then produced.  At $\vec{p}=\vec{0}$ a definite parity can be projected out using the operator
\begin{equation}\label{ParityProjector}
\Gamma_{\pm} = \frac{1}{2}\,(\gamma_{0} \pm I)\,.
\end{equation}
Defining $G_{ij}(\vec{p},\, t) = \Tr\left(\Gamma \, \mathcal{G}_{ij}(\vec{p},\, t)\right)$,  we
can write the Dirac-traced correlation function as a sum of exponentials,
\begin{equation}
\label{GijSum}
{G}_{ij}(t) = \sum_{\alpha}\lambda^{\alpha}_{i}\,\bar{\lambda}^{\alpha}_{j}\,\textrm{e}^{-m_{\alpha}t}.
\end{equation}
Here $\lambda^{\alpha}_{i}$ and $\bar{\lambda}^{\alpha}_{j}$ are the couplings of the annihilation, 
$\chi_{i}$, and creation, $\overline{\chi}_{j}$, operators at the sink and source respectively, while
the energy eigenstates of mass $m_{\alpha}$ are enumerated by $\alpha$.  We then search for a
linear combination of operators
\begin{equation}
\bar{\phi}^{\alpha} = \bar{\chi}_{j}\,u^{\alpha}_{j} \qquad \textrm{ and } \qquad \phi^{\alpha} = \chi_{i}\,v^{\alpha}_{i}
\end{equation}  
such that $\phi$ and $\bar{\phi}$ couple to a single energy eigenstate.
It is then clear from Eq.~(\ref{GijSum}) that 
\begin{equation}
{G}_{ij}(t_{0} + \dt)\,u^{\alpha}_{j} = \textrm{e}^{-m_{\alpha}\dt}\,{G}_{ij}(t_{0})\,u^{\alpha}_{j},
\end{equation}
and therefore for a given choice of variational parameters $(t_0,\dt)$, $u^{\alpha}_{j}$ and
$v^{\alpha}_{i}$ can be obtained by solving the left and right eigenvalue equations
\begin{align}
\label{E-value-eq-L}
\big[{G}^{-1}(t_{0})\,{G}(t_{0} + \dt)\big]_{ij}\,u^{\alpha}_{j} &= c^{\alpha}\,u^{\alpha}_{i}\\
\label{E-value-eq-R}
v^{\alpha}_{i}\,\big[{G}(t_{0} + \dt)\,{G}^{-1}(t_{0})\big]_{ij} &=
c^{\alpha}\,v^{\alpha}_{j},
\end{align}
with eigenvalue $c^{\alpha} = \textrm{e}^{-m_{\alpha}\dt}$.  We note that ${G}_{ij}$ is a symmetric
matrix in the ensemble average and as such we work with the improved estimator
$\frac{1}{2}\,({G}_{ij} + {G}_{ji})$ that ensures the eigenvalues of Eqs. (\ref{E-value-eq-L}) and
(\ref{E-value-eq-R}) are equal.  At $t_{0}$ and $t_{0} +\dt$ the eigenvectors $u^{\alpha}_{j}$ and
$v^{\alpha}_{i}$ diagonalise the correlation matrix which enables us to write down the
eigenstate-projected correlation function as
\begin{equation}
{G}^{\alpha}(t) = v^{\alpha}_{i}\,{G}_{ij}(t)\,u^{\alpha}_{j}.
\end{equation}
${G}^{\alpha}(t)$ is then used to extract masses.  Further details on the extraction of energy
eigenstates by performing a covariance matrix based $\chi^{2}$ analysis to fit the single state
Ansatz can be found in Ref.~\cite{Kiratidis:2015vpa}.  The analysis can also be performed on a
symmetric matrix with orthogonal eigenvectors, the details of which are found in
Ref.~\cite{Mahbub:2012ri}.  We now turn our attention toward the selection of interpolating
operators.

%
%%%%%%%%%%%%%%%%%%%%%%%%%%%%%%%%%%%%%%%%%%%%%%%
%
\section{Interpolating Fields}
\label{sect:InterpolatingFields}

As discussed in Section \ref{sect:Introduction}, previous work with five-quark
operators~\cite{Kiratidis:2015vpa, Lang:2012db} have successfully extracted states consistent with
scattering thresholds in the negative-parity nucleon channel.  Of particular note is
Ref.~\cite{Lang:2012db} in which a state consistent with the S-wave $\pi N$ scattering threshold
was extracted with a high degree of precision after explicitly projecting single-hadron momentum
onto each single-hadron factor of the five-quark operator.  Both of these calculations employed a
$\pi N$-type negative-parity five-quark operator, constructed with a negative-parity meson piece
and a positive-parity nucleon piece.  Motivated by this success, we investigate a similar tactic in
the positive-parity channel.  Here, we construct five-quark operators with a positive-parity meson
piece and a positive-parity nucleon piece.

Utilising the operators for the positive-parity isocsalar $\sigma$ and isovector $a^{0}_{0}$ and
$a^{+}_{0}$ mesons
\begin{align}
\sigma &= \frac{1}{\sqrt{2}}\big[\bar{u}^{e}\,I\,u^{e} + \bar{d}^{e}\,I\,d^{e}\big]\, , \nonumber\\
a^{0}_{0} &= \frac{1}{\sqrt{2}}\big[\bar{u}^{e}\,I\,u^{e} - \bar{d}^{e}\,I\,d^{e}\big]\, , \nonumber\\
a^{+}_{0} &= \big[\bar{d}^{e}\,I\, u^{e}\big]\, ,
\end{align}
we can construct five-quark $\sigma{N}$- and $a_{0}{N}$-type interpolators.  Recalling that the
$\sigma$ meson has the same quantum numbers as the vacuum we are able to write down the general
form of the $\sigma{N}$-type interpolators as
\begin{align}
\chi_{\sigma{N}}(x) &= \frac{1}{2}\,\epsilon^{abc}\,\big[u^{Ta}(x)\,\Gamma_{1}\,d^{b}(x)\big]\,\Gamma_{2}\,u^{c}(x)\nonumber\\
&\qquad\qquad\times\big[\bar{u}^{e}(x)\,I\,u^{e}(x) + \bar{d}^{e}(x)\,I\,d^{e}(x)\big].
\end{align}
Here, the choices of $(\Gamma_{1}, \Gamma_{2}) = (C\gamma_{5}, \textrm{I})$ and $(C,\gamma_{5})$
provide us with two five-quark operators $\chi_{\sigma{N}}(x)$ and $\chi^{\prime}_{\sigma{N}}(x)$
respectively.

Similarly, we write down the general form of the $a_{0}{N}$-type operators using the Clebsch-Gordan
coefficients to project isospin $I = 1/2, I_{3} = +1/2$ obtaining
\begin{align}\label{Proton5QrkOpInterpolator}
\chi_{a_0{N}}(x) &= \frac{1}{\sqrt{6}}\,\epsilon^{abc}\times\nonumber\\
&\quad\bigg\{2\big[u^{Ta}(x)\,\Gamma_{1}\,d^{b}(x)\big]\,\Gamma_{2}\,d^{c}(x)\,\big[\bar{d}^{e}(x)\,I\,u^{e}(x)\big]\nonumber\\
&\quad - \big[u^{Ta}(x)\,\Gamma_{1}\,d^{b}(x)\big]\,\Gamma_{2}\,u^{c}(x)\,\big[\bar{d}^{e}(x)\,I\,d^{e}(x)\,\big]\nonumber\\
&\quad + \big[u^{Ta}(x)\,\Gamma_{1}\,d^{b}(x)\big]\,\Gamma_{2}\,u^{c}(x)\,\big[\,\bar{u}(x)^{e}\,I\,u^{e}(x)\big]\bigg\},
\end{align}
where the two aforementioned choices of $(\Gamma_{1}, \Gamma_{2})$ provide $\chi_{a_0{N}}(x)$ and
$\chi^{\prime}_{a_0{N}}(x)$ respectively.  In addition, we include the two
five-quark operators $\chi_{\pi{N}}$ and $\chi^{\prime}_{\pi{N}}$ based on the form
\begin{align}\label{Proton5QrkOpInterpolator}
\chi_{\pi{N}}(x) &= \frac{1}{\sqrt{6}}\,\epsilon^{abc}\,\gamma_{5}\times\nonumber\\
&\quad\bigg\{2\big[u^{Ta}(x)\,\Gamma_{1}\,d^{b}(x)\big]\,\Gamma_{2}\,d^{c}(x)\,\big[\bar{d}^{e}(x)\,\gamma_{5}\,u^{e}(x)\big]\nonumber\\
&\quad - \big[u^{Ta}(x)\,\Gamma_{1}\,d^{b}(x)\big]\,\Gamma_{2}\,u^{c}(x)\,\big[\bar{d}^{e}(x)\,\gamma_{5}\,d^{e}(x)\,\big]\nonumber\\
&\quad + \big[u^{Ta}(x)\,\Gamma_{1}\,d^{b}(x)\big]\,\Gamma_{2}\,u^{c}(x)\,\big[\,\bar{u}(x)^{e}\,\gamma_{5}\,u^{e}(x)\big]\bigg\},
\end{align}
and detailed in Ref.~\cite{Kiratidis:2015vpa}.  Our basis of qualitatively different operators is
completed with the inclusion of the standard three-quark nucleon operators $\chi_{1}$ and
$\chi_{2}$ given by
\begin{align}\label{NucleonInterps}
\chi_{1} &= \epsilon^{abc}[u^{aT}\, (C\gamma_{5})\, d^{b}]\, u^{c}\nonumber\\
\chi_{2} &= \epsilon^{abc}[u^{aT}\, (C)\, d^{b}]\, \gamma_{5}\, u^{c}.
\end{align}

With the introduction of five-quark operators having an anti-quark flavour matching one of the
quark flavours, diagrams that contain loop propagators $S(y,y)$ where the source and sink position coincide are encountered.  Loops at the source, $S(0,0)$, can be treated with conventional propagators via $S(x,0)|_{x=0}$.  Loops at
the sink, $S(x,x)$, are stochastically estimated by averaging over four independent
$\mathbb{Z}_4$ noise vectors that are fully diluted in time, spin and colour.  Further details of
our calculation of these stochastic propagators, along with the method by which they are smeared, 
can be found in our previous work \cite{Kiratidis:2015vpa}.

\section{Simulation Details}
\label{sect:Simulation Details}

The results presented herein utilise the PACS-CS $2 + 1$ flavour dynamical-fermion
configurations~\cite{Aoki:2008sm} with an Iwasaki gauge action~\cite{Iwasaki:1983ck} which are made
available through the ILDG~\cite{Beckett:2009cb}.  The lattice size is $32^3 \times 64$ with a
lattice spacing of $0.0907\textrm{ fm}$ which provides a physical volume of $\approx (2.90\textrm{
  fm})^3$.  The light quark mass is set by the hopping parameter $\kappa_{ud} = 0.13754$ which
gives a pion mass of $m_{\pi} = 411 \textrm{ MeV}$, while the strange quark mass is set by
$\kappa_{s} = 0.13640$.  Backward propagating states are removed via the imposition of fixed
boundary conditions in the time direction~\cite{Melnitchouk:2002eg, Mahbub:2009nr}.

The source insertion occurs at time slice $t_{src} = n_t/4 = 16$, well away from the 
boundary and its associated effects.  Our variational analysis is performed with parameters $(t_{0},\dt)=(17,3)$
which provides a good balance between systematic and statistical uncertainties.  Error bars are
calculated via single elimination jackknife, while a full covariance matrix analysis provides the
$\chi^{2}/dof$, which is used to select appropriate fit regions for eigenstate-projected
correlators.

Gauge-invariant Gaussian smearing~\cite{Gusken:1989qx} at the source and sink is used to increase
the span of our basis by altering the overlap of our interpolators with the states of interest.  We
investigate three levels of $n_s = 35, 100$ and 200 sweeps of Gaussian smearing.

%%%%%%%%%%%%%%%%%%%%%%%%%%%%%%%%%%%%%%%%%%%%%%%%%%%%%%%%%%%%%
%
\section{Results}
\label{sect:Results}

\subsection{Correlation Matrix Construction}

As we now possess eight qualitatively different operators, each with three different levels of
Gaussian smearing, our basis of twenty-four operators admits a substantial number of possible
sub-bases of interest.  Consequently, it is instructive to investigate various ratios of
correlators, in order to determine which combinations can provide suitable sub-bases such that the
condition number of the correlation matrix is favorable.

\begin{figure}[!htbp]
  \centering 
  {\includegraphics[width=0.48\textwidth]{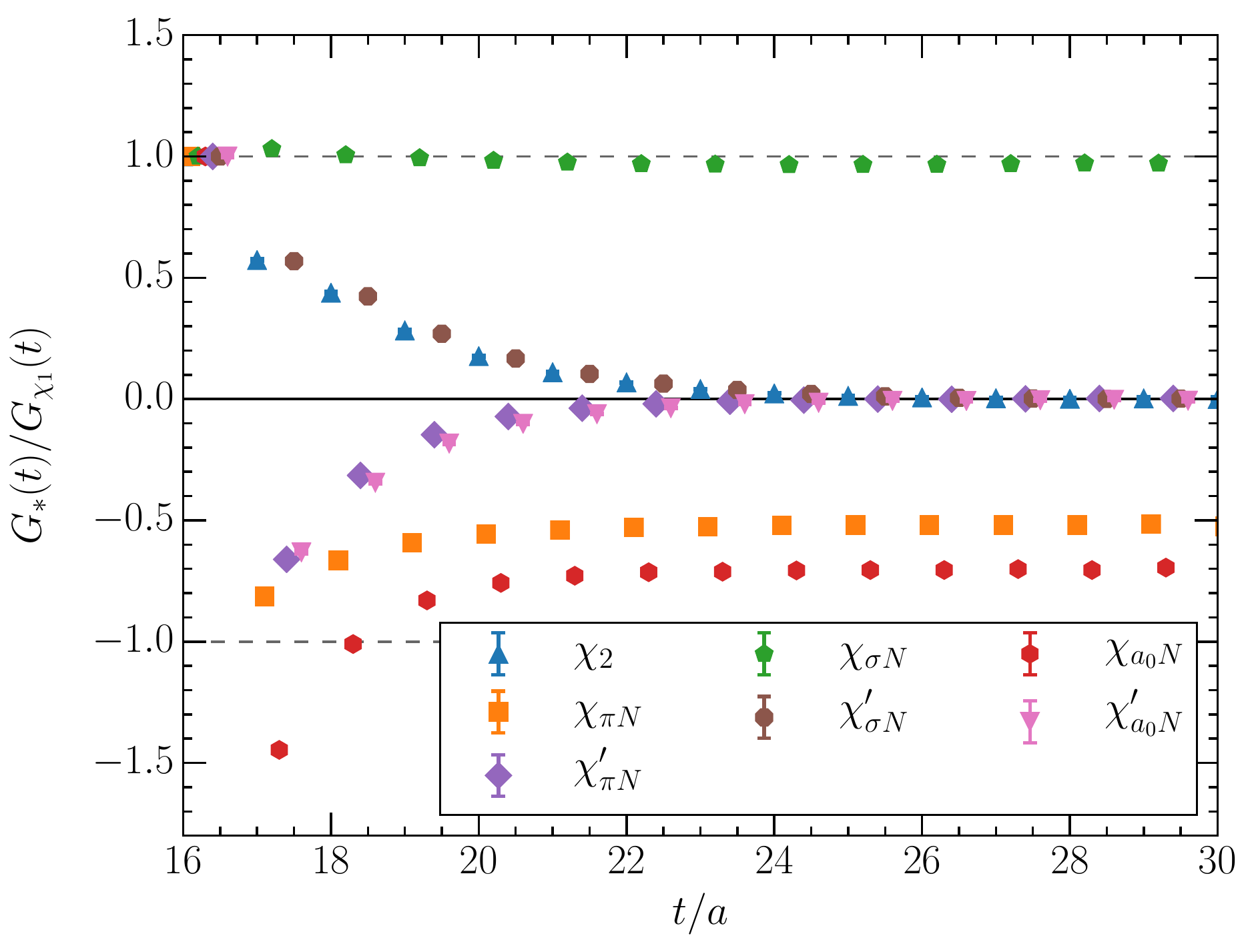}}\\
  {\includegraphics[width=0.48\textwidth]{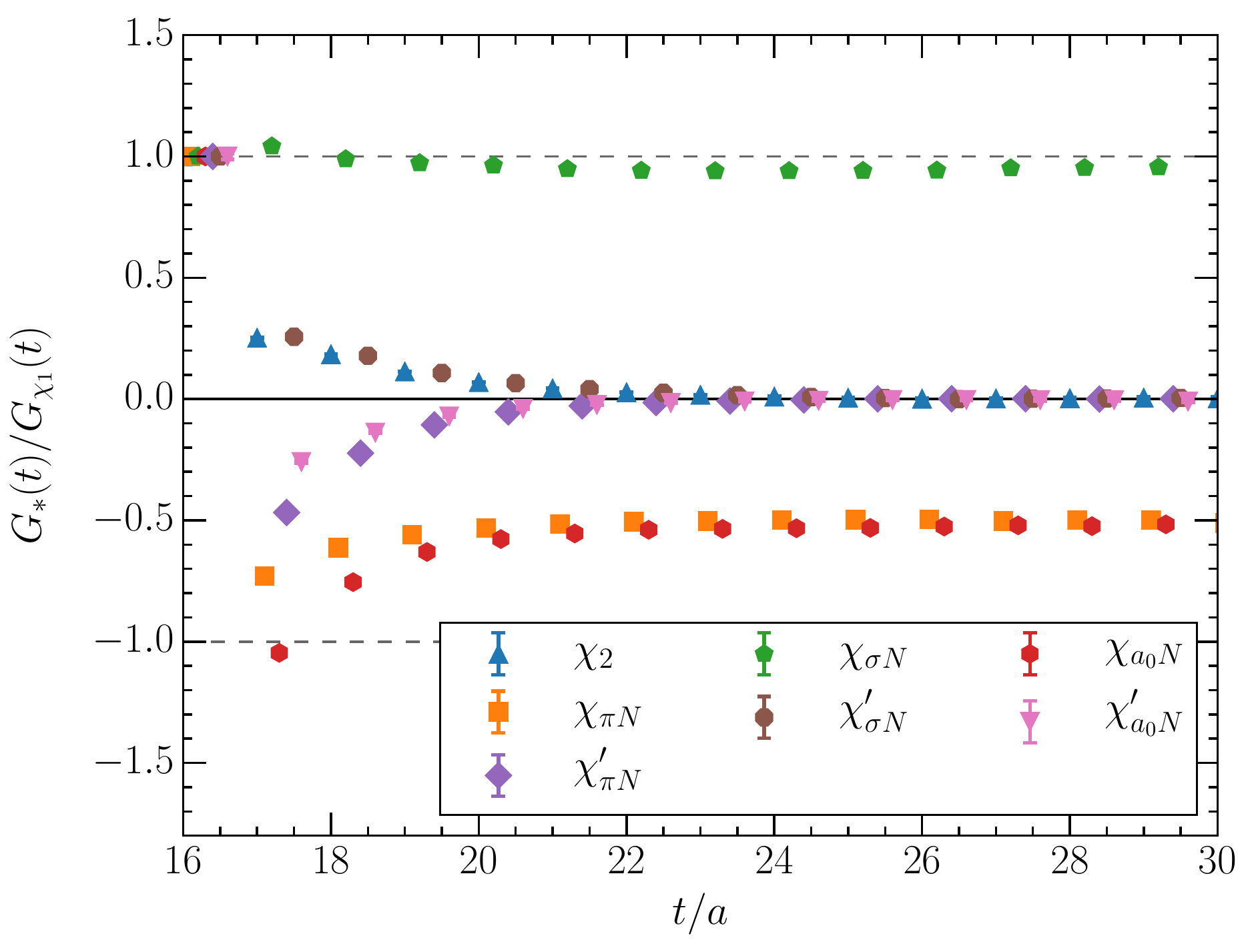}}\\
  {\includegraphics[width=0.48\textwidth]{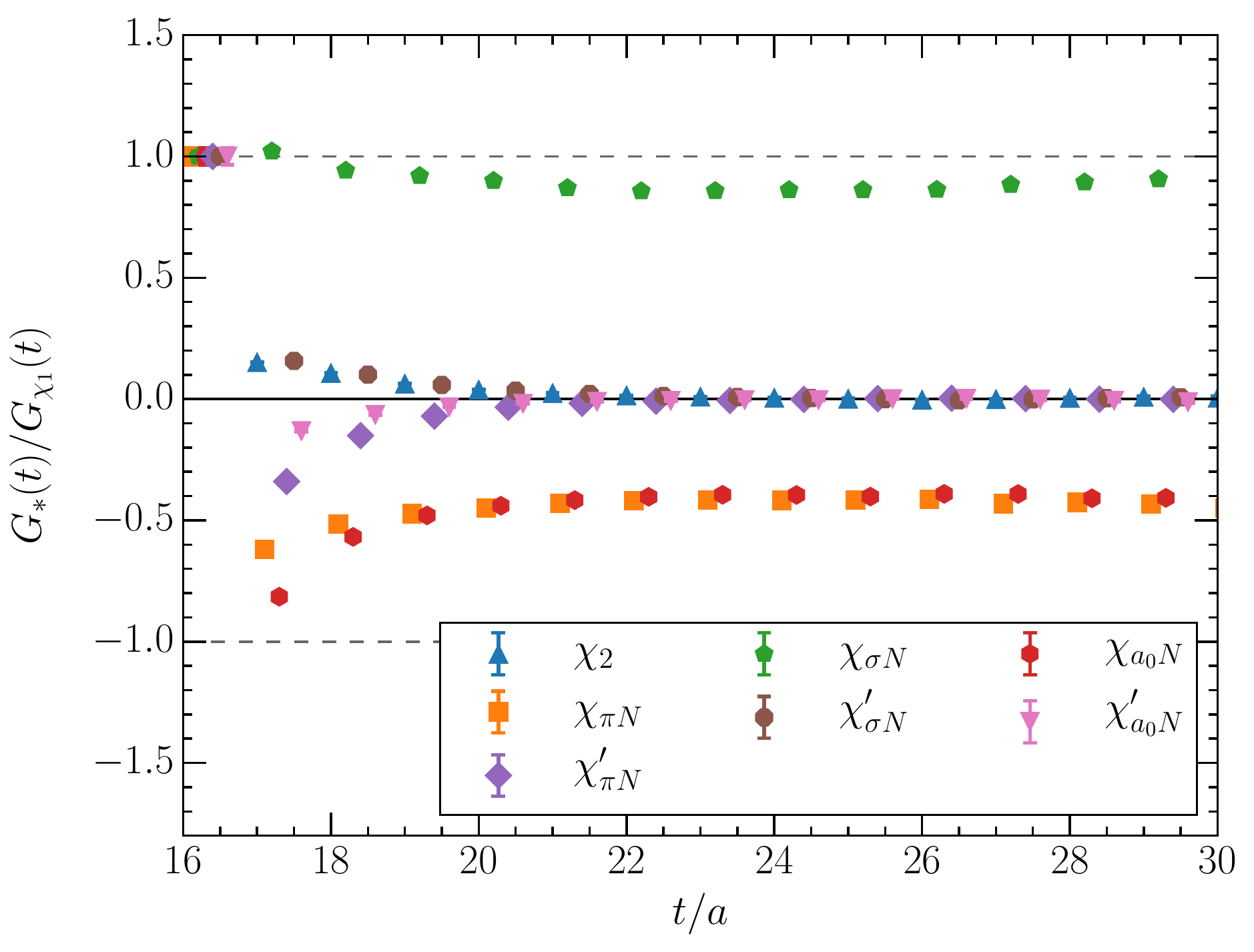}}
\caption{(Colour online).  Correlation function ratios constructed to illustrate different
  superpositions of energy eigenstates in the correlators.  The ratio is formed by dividing the
  correlator corresponding to each operator indicated in the legend by the correlation function
  formed from the $\chi_{1}$ operator.  Plots are presented at 35 (top), 100 (middle) and 200
  (bottom) sweeps of Gaussian smearing in the quark-propagator source and sink.  For clarity of
  presentation, the $t$ component of the ratio is sequentially offset.}
\label{fig:CorrelatorRatioComparison}
\end{figure}

In Figure \ref{fig:CorrelatorRatioComparison} we present plots at each of the three smearing levels
studied, showing a ratio of correlators formed by dividing each correlator with the correlation function
formed from the standard $\chi_{1}$ operator.  Our aim is to identify correlators showing a unique
approach to the plateau, indicating a novel superposition of excited states.

Notably, the ratios formed from the $\sigma{N}$ type operators, that is
${G}_{\chi_{\sigma{N}}}/{G}_{\chi_{1}}$ and ${G}_{\chi^{\prime}_{\sigma{N}}}/{G}_{\chi_{1}}$ behave
in a remarkably similar manner to the ratios ${G}_{\chi_{1}}/{G}_{\chi_{1}}$ and
${G}_{\chi_{2}}/{G}_{\chi_{1}}$.
Consequently, we anticipate the overlap of $\chi_{\sigma{N}}$ with states in the spectrum is very
similar to $\chi_{1}$ and similarly the overlap of states with $\chi^{\prime}_{\sigma{N}}$ is much
the same as with $\chi_{2}$.  Evidently, these new $\sigma{N}$-type operators provide little new
information from that already contained in $\chi_{1}$ and $\chi_{2}$.  Recalling that the $\sigma$
meson carries the quantum numbers of the vacuum provides some insight into this result.  In light
of this similarity, the $\chi_{\sigma{N}}$ and $\chi^{\prime}_{\sigma{N}}$ interpolators are
omitted from bases that also contain the matching $\chi_{1}$ or $\chi_{2}$ interpolator.

Of the two new $a_{0}{N}$ interpolators, $\chi_{a_{0}{N}}$ stands out from the other interpolators
at all three smearing levels.  $\chi_{a_{0}{N}}$ excites a novel superposition of nucleon
excited states and will aid in spanning the space of low-lying states.  It holds promise to reveal
the presence of a low-lying state missed in previous analyses.
Similarly, at 100 and especially 200 smearing sweeps, $\chi^{\prime}_{a_{0}{N}}$ shows a unique
path to the plateau, again indicating the promise of disclosing new states.

In comparing the various smearing levels for all the correlator ratios presented, one observes that the
plateau in the ratios occurs at earlier times as the smearing increases.  Again, these differences
between different smearing levels aid in spanning the space and generating correlation matrices
with good condition numbers.  However, the construction of large correlation matrices tends to increase the condition number and
decrease the likelihood of obtaining a solution.  This effect, combined with the larger
statistical uncertainties encountered with the largest smearing extent, leads to difficulties in finding
a solution to the generalised eigenvalue equations with the new five-quark operators.

\begin{table}[t]
\caption{The interpolating fields used in constructing each correlation-matrix basis.}
\begin{ruledtabular}
\begin{tabular}{cl}
    Basis Number & Operators Used  \\[2pt]
    \hline 
\noalign{\vspace{2pt}}
    1 & $\chi_{1}$, $\chi_{2}$\\
    2 & $\chi_{1}$, $\chi_{2}$, $\chi_{a_0{N}}$\\
    3 & $\chi_{1}$, $\chi_{2}$, $\chi_{a_0{N}}$, $\chi^{\prime}_{a_0{N}}$\\
    4 & $\chi_{\pi{N}}$, $\chi^{\prime}_{\pi{N}}$, $\chi_{a_0{N}}$\\
    5 & $\chi_{\pi{N}}$, $\chi^{\prime}_{\pi{N}}$, $\chi_{a_0{N}}$, $\chi^{\prime}_{a_0{N}}$\\
    6 & $\chi_{\pi{N}}$, $\chi^{\prime}_{\pi{N}}$, $\chi_{\sigma{N}}$, $\chi^{\prime}_{\sigma{N}}$\\
    7 & $\chi_{\sigma{N}}$, $\chi^{\prime}_{\sigma{N}}$, $\chi_{a_0{N}}$, $\chi^{\prime}_{a_0{N}}$\qquad\null\\[2pt]
\end{tabular}
\end{ruledtabular}
\label{table:BasisTable}
\end{table}

As a result, we focus on correlation matrices formed from 35 and 100 sweeps of smearing in the
propagator sources and sinks.  These are the smearing levels that provide the most variation at
early times, and hence the levels at which we are able to construct bases more likely to provide an
effective span of the state space, particularly in comparison to that obtained using three-quark
operators alone.  This enables us to examine scenarios with multiple, qualitatively different quark
structures, while still retaining the presence of multiple smearing levels.  
While we will not detail the results including the 200 sweep correlators, we do note that when a
solution was found, the energies of the low-lying eigenstates agreed with the results 
presented in the following.

We consider seven different correlation matrices formed from the bases outlined in Table
\ref{table:BasisTable}.  Each basis is formed with 35 and 100 sweeps of smearing, thus creating
four 8 $\times$ 8 bases, two 6 $\times$ 6 bases and one 4 $\times$ 4 basis.  While each correlation
matrix may disclose different states, the energies of the states observed should agree among the
different bases considered.

\subsection{Finite Volume Spectrum of States}

The development of Hamiltonian effective field theory
\cite{Hall:2013qba,Hall:2014uca,Liu:2016uzk,Liu:2016uzk} can provide some insight into the spectrum
to be anticipated.  By using an effective field theory model constrained to the experimental phase
shifts, inelasticities and pole position, one can predict the spectrum to be observed in the finite
volume of the lattice.  In Ref.~\cite{Liu:2016uzk}, three models are considered with different
roles played by the bare basis-state in constructing the Hamiltonian model.  In the popular model
incorporating a bare basis state with a mass of 2.0 GeV, the model predicts a Roper-like state at
1750 MeV in the finite volume of the lattice for the quark mass corresponding to $\kappa=0.13754$ that is considered herein.

Alternatively,
the third model of Ref.~\cite{Liu:2016uzk}, preferred by previous lattice results, predicts the
absence of low-lying states with a strong bare-state component, predicting instead the existence of
five meson-baryon scattering states below the state observed in lattice QCD, commencing at 1600
MeV.  Attraction in these channels could localize the meson to the vicinity of the baryon
\cite{Liu:2016wxq}, overcoming the volume suppression of the coupling.

\begin{figure}[t]
  \centering 
  {\includegraphics[width=0.48\textwidth]{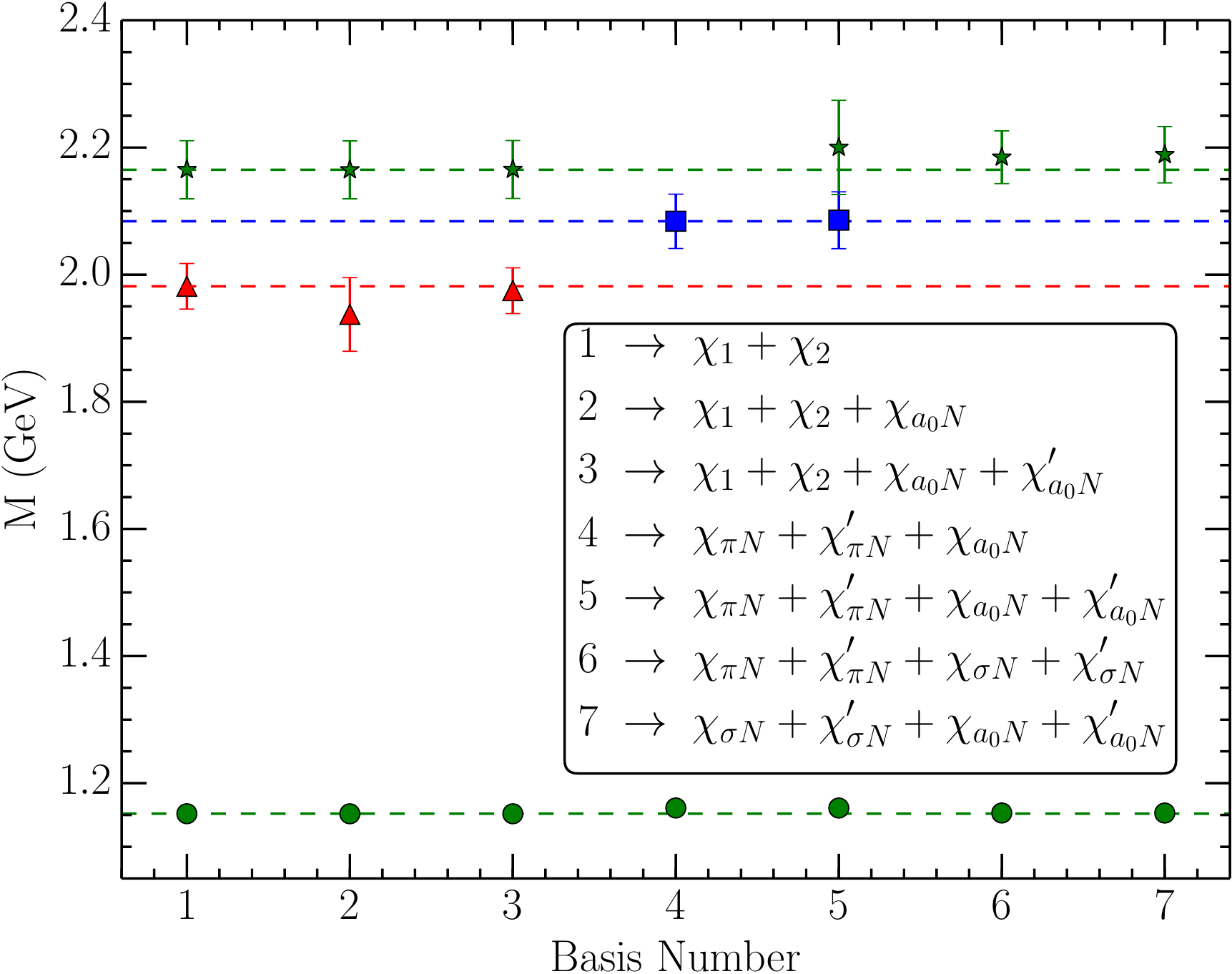}}\\
\caption{(Colour online).  Low-lying states observed for each of the correlation-matrix bases
  described in Table \ref{table:BasisTable}.  For each interpolating field, two smearing levels of
  $n_{s} = 35$ and $n_{s} = 100$ are used in all cases.  Dashed horizontal lines are present to guide the eye.  They have been set by the central values from basis 1 in all cases except for the state $\sim 2.1$ GeV, in which case it is drawn from basis 4.}
\label{fig:SpectraPlot}
\end{figure}

The low-lying spectra produced from the correlation matrices detailed in Table
\ref{table:BasisTable} are presented in Figure \ref{fig:SpectraPlot}.
In basis number one, we present results from a simple 4 $\times$ 4 analysis with the three-quark
$\chi_{1}$ and $\chi_{2}$ interpolating fields at two different smearing levels.  This
consideration of three-quark operators alone \cite{Mahbub:2009aa,Mahbub:2010rm} provides the
benchmark analysis that we will refer to as we attempt to ascertain whether subsequent bases with
five-quark operators alter the low-lying spectrum.  

As previously mentioned, $\chi_{a_{0}N}$ appeared to be the most promising new operator, in that
the ratio $\chi_{a_{0}N}/\chi_{1}$ displayed the largest variation when compared to ratios
previously studied.  As such, in column two we add the $\chi_{a_{0}N}$ operator to $\chi_{1}$ and
$\chi_{2}$ and perform the resulting 6 $\times$ 6 correlation matrix analysis.  This analysis
reveals no new low-lying states.  We then further add the $\chi^{\prime}_{a_{0}N}$ operator forming
an 8 $\times$ 8 analysis.  Once again the spectrum is invariant, revealing no new states.

As the overlap of three-quark operators with three-quark states is naturally large when compared to
five-quark operators, we proceed by considering bases that contain only five-quark operators.  The
aim is to allow spectral strength that may have ordinarily been overwhelmed by three-quark
operators to come to the fore.  Such an approach was beneficial in the odd-parity nucleon sector
\cite{Kiratidis:2015vpa}.

The results of a $6 \times 6$ analysis using $\chi_{\pi{N}}$, $\chi^{\prime}_{\pi{N}}$ and
$\chi_{a_{0}N}$ are illustrated as basis number four in Fig.~\ref{fig:SpectraPlot}.  Here we do
observe a state between the two previously observed states, but crucially no new low-lying state is
extracted.  To ascertain whether the observed state is new, or a superposition of the two states
observed previously, we consider larger five-quark operator bases.

In the final three columns we form $8 \times 8$ bases with the three possible combinations of pairs
of our five-quark operators.  States consistent with those already observed are extracted,
including the new state observed in basis four.  However, no new low-lying states are found.

Returning to the aforementioned Hamiltonian effective field theory model \cite{Liu:2016uzk}, there
are some common features in the spectrum.
The splitting of $\sim 200$ MeV  between the first and second
excitations observed with the three-quark operators is similar to that predicted by the model.
More  interesting is the model's prediction of a scattering state
with a dominant $\pi{N}$ component roughly half way between the two excitations.
In bases four and five, containing $\chi_{\pi{N}}$ and $\chi^{\prime}_{\pi{N}}$ interpolators, we
do observe a new state roughly half way between the two excitations.  The dismissal of three-quark
operators is key to disclosing this state.

While these qualitative features are consistent, the goal of this investigation was to reveal new
states below the lowest-lying excitation of three quark operators through the consideration of
novel five-quark operators.
We are now able to conclude that the introduction of positive-parity mesons in local five-quark
operators of the nucleon, does not provide strong overlap with the anticipated low-lying finite-volume
scattering states.  

However, these operators do have strong overlap with the ground state nucleon, once again
highlighting the meson-baryon cloud of the nucleon.  In bases four through seven, only five-quark
operators are considered and we are able to extract the ground state mass with a high degree of
precision, comparable to that obtained solely with three-quark operators.

\section{Conclusions}
\label{sect:Conclusions}

In this exploratory investigation we have introduced local five-quark 
operators with the quantum numbers of the Roper resonance,
based on combining positive-parity mesons with conventional nucleon interpolators.  Drawing on
success in the negative parity channel, the aim was to reveal new low-lying states that had been
missed in previous calculations utilizing three-quark operators.  The construction of $a_{0}{N}$-
and $\sigma{N}$-type interpolating fields was outlined and variational analyses were performed with
these new interpolating fields, in combination with previously considered $\pi{N}$-type and
standard three-quark interpolators.

Ratios of correlation functions were examined to discover which interpolators gave rise to new
superpositions of excited states and therefore which interpolators held the greatest promise of
overlapping with new states.  This process indicated that the $\chi_{a_{0}{N}}$ operator was the
most promising interpolator for revealing new low-lying states.

Correlation matrices were constructed from several different bases of interpolating operators.  By systematically varying the operators used, 
the independence of the low-lying spectrum from the basis
could be checked and the potential for new state discovery was increased.  In accord with previous
studies, changing the operators composing the basis of a correlation matrix does affect whether or
not a particular state is observed.  

While a new state anticipated by Hamiltonian effective field theory was observed in this analysis,
no new states below the first excitation found with three-quark operators were observed.  The local
five-quark operators studied were found to posses a strong overlap with the ground state nucleon,
as bases containing only these operators produced a ground state with a high degree of precision.

We conclude that the low-lying finite-volume meson-baryon scattering states anticipated by
Hamiltonian effective field theory are not well localised.  Instead, the states appear to be weakly
interacting such that the volume suppression of two-particle scattering states with local operators
prevents their strong overlap with the interpolators considered herein.  The results strengthen the
interpretation of the Roper as a coupled-channel dynamically-generated meson-baryon resonance, a
resonance not closely associated with conventional three-quark states.

\section*{Acknowledgments}
We thank the PACS-CS Collaboration for making these $2+1$ flavor configurations available and the
ongoing support of the ILDG.  The majority of these calculations were performed on the University
of Adelaide's Phoenix cluster.  This research was undertaken with the assistance of resources at
the NCI National Facility in Canberra, Australia.  These resources were provided through the
National Computational Merit Allocation Scheme, supported by the Australian Government and the
University of Adelaide Partner Share.  This research is supported by the Australian Research
Council through the ARC Centre of Excellence for Particle Physics at the Terascale (CE110001104),
and through Grants No.\ LE160100051, DP151103101 (A.W.T.), DP150103164, DP120104627 and LE120100181
(D.B.L.).

\bibliographystyle{apsrev4-1}
\bibliography{reference}

\end{document}